# THE INFRARED SENSITIVITY OF NONABELIAN DEBYE SCREENING[*]

A. K. REBHAN

*DESY, Gruppe Theorie,*
*Notkestr. 85, D-22603 Hamburg, Germany*

*In memory of Tanguy Altherr*

ABSTRACT

It is shown that a perturbative treatment of nonabelian Debye screening at high temperature suffers from infrared problems already at the next-to-leading order, which is given by ring-resummed one-loop diagrams. Superficial infrared power counting would let one expect a sensitivity to the magnetic mass scale only at higher orders, but the form of Debye screening depends on the analytic structure of the correlation functions, which is strongly sensitive to the existence of screening of static magnetic fields.

## 1. Introduction

High-temperature gauge theories[1] become increasingly complicated as the infrared regime is approached. At the momentum scale $gT$, where $g$ is the coupling and $T$ the temperature, the spectrum of the theory begins to deviate considerably from the one of the free theory. Gauge bosons as well as fermions acquire new, collective degrees of freedom described by the gauge-invariant (nonlocal) effective action generated by the so-called "hard thermal loops" (HTL)[2,3], the high-temperature limit of all one-loop Feynman diagrams proportional to $T^2$. The resulting leading-order dispersion laws of the quasi-particle excitations can be understood in classical terms[4,5], but already at next-to-leading order the dispersion laws receive contributions from all orders of the conventional perturbation series. In order to restore perturbation theory it is necessary to resum all HTL corrections[3], and in this way some corrections to the HTL dispersion laws have been obtained[6,7,8]. However, in nonabelian theories, the restitution of perturbation theory is not complete. Static chromomagnetic fields are not screened at the HTL level, and the self-interactions of these lead to a breakdown of perturbation theory at a certain loop order. In particular, the magnetic screening mass, which is expected to arise at the order $g^2 T$, is not calculable in perturbation theory[9,10]; HTL resummation does not help in this respect.

---

[*]Invited talk at the Workshop on Quantum Infrared Physics, 6–10 June 1994, American University of Paris, France.

On the other hand, HTL resummation should allow one to calculate corrections to the classical chromoelectric (Debye) screening mass $m_0$, which is of the order $gT$, to wit,

$$m_0^2 = \frac{e^2 T^2}{3} \qquad (1)$$

with $e^2 = (N + N_f/2)g^2$ for color group SU($N$) and $N_f$ flavors. Upon HTL resummation, perturbation theory is organised in powers of $g$ rather than $g^2$, and the relative order $g$ correction is determined by resummed one-loop diagrams. According to superficial infrared power-counting[1], problems coming from the perturbatively vanishing magnetic mass would be expected to set in at two-loop order.

In this note, I shall demonstrate, however, that already the next-to-leading order term of the resummed perturbation theory is crucially sensitive to the magnetic mass scale.

## 2. The electrostatic gluon propagator at resummed one-loop order

In momentum space, the 'electrostatic' gluon propagator is given by

$$D_L(k) = \frac{1}{k^2 + \Pi_{00}(k_0 = 0, k)} \qquad (2)$$

and its Fourier transform $\Phi(r)$ determines the chromoelectric field induced by a single external (conserved) source $J$ through

$$\langle E^i(\mathbf{x}) \rangle = \frac{\partial}{\partial x_i} \Phi(|\mathbf{x}|), \qquad (3)$$

where we have suppressed all color indices. Despite the nonabelian nature, this is an exact result of linear response theory[11], since with only one direction in color space introduced by the single source $J$, the gauge potentials all point in the same direction, which eliminates the nonabelian commutator terms trivially.

At high temperature, the leading contribution to $\Pi_{00}(k_0 = 0, k)$ equals the constant $m_0^2$ of Eq. (1), and $\Phi(r) \propto e^{-m_0 r}/r$. However, beyond leading order $\Pi_{00}$ is a gauge fixing dependent quanitity. At one-loop resummed order one obtains[12,13] in covariant gauges with gauge parameter $\alpha$

$$\Pi_{00}(0, k) = m_0^2 + \frac{g\sqrt{3N'}m_0^2}{2\pi}\left[\frac{m_0^2 - k^2}{m_0 k}\arctan\frac{k}{m_0} + \frac{\alpha - 2}{2}\right] + O(g^2), \qquad (4)$$

where $N' = N/(1 + \frac{N_f}{2N})$. Computing the Fourier transform of $D_L(k)$ with the thus corrected gluon self-energy, one finds a surprising behavior[14]: $\Phi(r)$ has the form of a Yukawa potential multiplied by an oscillating function which approaches a negative value asymptotically. Its details are gauge parameter dependent except for the screening mass which characterises the exponential decay for very large distances, which is still given by $m_0$.

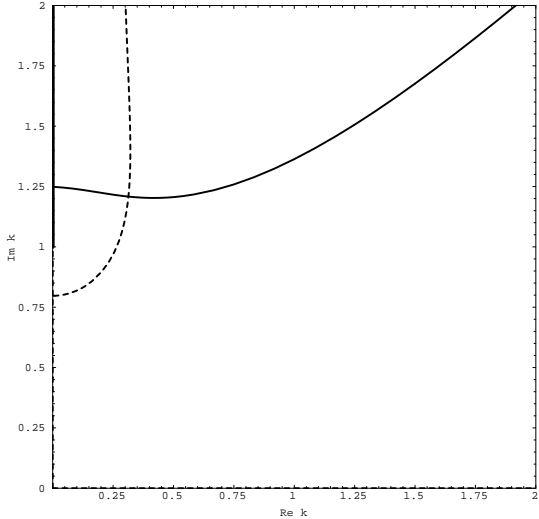

Figure 1: The analytic structure of $D_L^{-1} \equiv k^2 + \Pi_{00}(0,k)$ in units where $1 = m_0$ and with coupling $t \equiv g\sqrt{3N'}/8 = 0.25$. The full line corresponds to $\Re e D_L^{-1} = 0$ and the dashed one to $\Im m D_L^{-1} = 0$. The latter extends to all of the real and imaginary axes except for the thick part of the imaginary axis which marks the location of the branch cut. The intersection of the full and the dashed line corresponds to a pole of $D_L$ and is located at $k/m_0 \approx 1.208 + 0.313i$.

The reason behind this peculiar result is displayed in Fig. 1, where the analytic structure of $D_L^{-1}(k)$ is rendered for $t \equiv g\sqrt{3N'}/8 = 0.25$ and gauge parameter $\alpha = 1$. Shown is the first quadrant of the complex $k$ plane; the others are given by reflection on the real and imaginary axes. At leading order, $D_L(k)$ was determined by a simple pole at $k/m_0 = i$, but including the next-to-leading order, $D_L^{-1}$ no longer has a simple zero there, but a logarithmic branch singularity followed by a branch cut from $i$ to $\infty$. The original zero of $D_L^{-1}$, however, still exists: it has moved to the right (and also left) of the imaginary axis. With the above parameters, the corresponding poles of $D_L$ contribute to $\Phi$ a term proportional to $\cos(0.313x)e^{-1.208x}/x$, where $x \equiv rm_0$. There is, however, also the contribution from the cut, which adds a term $-f(x)e^{-x}/x$ with a strictly positive function $f$, so that asymptotically, for very large $x$, the behavior is that of a repulsive Yukawa potential with screening mass $m_0$.

However, the singularity of the next-to-leading correction to $\Pi_{00}(0,k)$ at $k = \pm im_0$ indicates a breakdown of perturbation theory in the vicinity of these points in momentum space, for there the correction term becomes larger than the supposed leading one. In Ref. [13] I have pointed out that this singularity might be cut off by a nonvanishing magnetic mass. Indeed, introducing a small magnetic mass changes $\Pi_{00}(0,k)$ significantly around $k = \pm im_0$. Denoting the magnetic mass by $m_m$ and

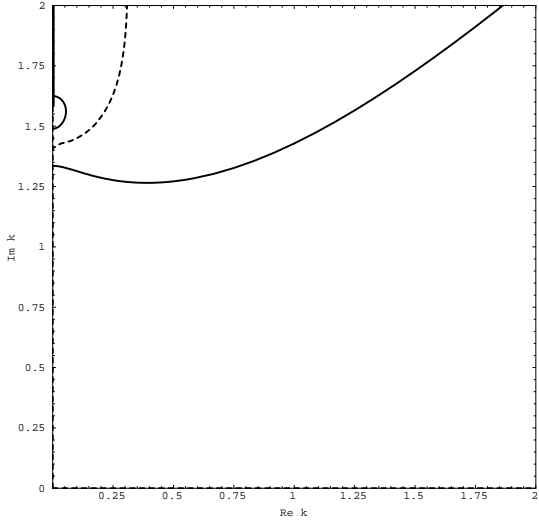

Figure 2: As in Fig. 1, but with a finite magnetic mass $m_m = 0.25m$.

the full Debye mass by $m$, $\Pi_{00}$ becomes

$$\Pi_{00}(0,k) = m_0^2 + \frac{g\sqrt{3N'}m_0}{2\pi}\Big[-\frac{1}{2}m - \frac{1}{2}m_m \qquad (5)$$
$$+ \frac{1}{k}(m^2 - \frac{1}{2}m_m^2 - k^2)\arctan\frac{k}{m+m_m}$$
$$+ \left(k^2 + m^2\right)\Big\{\frac{k^2+m^2}{m_m^2 k}\Big(\arctan\frac{k}{m+\sqrt{\alpha}m_m}$$
$$- \arctan\frac{k}{m+m_m}\Big) + (\sqrt{\alpha} - 1)\frac{1}{m_m}\Big\}\Big],$$

where $m_m$ has been introduced in a gauge-invariant manner by mimicking the Higgs mechanism[15]. Since the branch singularity now has moved to $k = \pm i(m+m_m)$, one can use a self-consistent definition of the Debye mass through

$$m^2 = \Pi_{00}(0,k)\Big|_{k^2=-m^2}, \qquad (6)$$

which restitutes a pole at purely imaginary $k = \pm im$. From (6) one finds that the mass $m$ thus defined is gauge parameter independent, and is given by $m^2 = m_0^2(1+\delta)$ with[13]

$$\delta \equiv \frac{\delta m^2}{m_0^2} = \frac{\sqrt{3N'}g}{2\pi}\left(\ln\frac{m}{m_m} + \ln 2 - \frac{1}{2}\right) + O(g^2). \qquad (7)$$

The full analytic structure of $D_L^{-1}(k)$ following from (6) is given in Fig. 2 for the same parameters as in Fig. 1, but now with nonvanishing $m_m/m = 0.25$. The lines $\Re e D_L = 0$ and $\Im m D_L = 0$ no longer intersect at complex values of $k$, but only on the imaginary axis. There is also a gauge dependent zero very close to the branch singularity, but the dominant singularity in $D_L$ is the one at $k = im$. Thus $\Phi(r)$ no longer changes sign but decays exponentially for large $x$ with gauge independent screening mass $m$.

I would like to thank H. Fried, B. Müller, and M. LeBellac for organizing this stimulating workshop, and R. Baier for continuous discussions. This research is supported in part by the EEC Programme "Human Capital and Mobility", Network "Physics at High Energy Colliders", contract CHRX-CT93-0357 (DG 12 COMA).